\documentclass[10pt,aps,prb,twocolumn, nofootinbib]{revtex4-1}

\usepackage{amssymb,amsmath}
\usepackage{subfig}
\usepackage{graphicx}
\usepackage{color}
\usepackage{hyperref}
\usepackage{listings}
\hypersetup{
  bookmarks=true,     
  unicode=false,     
  pdftoolbar=true,    
  pdfmenubar=true,    
  pdffitwindow=false,   
  pdfstartview={FitH},  
  pdfauthor={Alyssa M Adams},   
  colorlinks=true,    
  linkcolor=blue,     
  citecolor=red,    
}
\newtheorem{mydef}{Definition}

\begin{document}

\definecolor{dkgreen}{rgb}{0,0.6,0}
\definecolor{gray}{rgb}{0.5,0.5,0.5}
\definecolor{mauve}{rgb}{0.58,0,0.82}

\lstset{frame=tb,
 	language=Matlab,
 	aboveskip=3mm,
 	belowskip=3mm,
 	showstringspaces=false,
 	columns=flexible,
 	basicstyle={\small\ttfamily},
 	numbers=none,
 	numberstyle=\tiny\color{gray},
 	keywordstyle=\color{blue},
	commentstyle=\color{dkgreen},
 	stringstyle=\color{mauve},
 	breaklines=true,
 	breakatwhitespace=true
 	tabsize=3
}

\title{The Effects of Interaction Functions Between Two Cellular Automata}
\author{Alyssa M Adams}
\affiliation{Department of Bacteriology \& Computation and Informatics in Biology and Medicine Program, University of Wisconsin-Madison, Madison, WI USA \\ Algorithmic Nature Group \\ alyssa.gp.adams@gmail.com}
\date{\today}

\begin{abstract}

Biological systems are notorious for complex behavior within short timescales (e.g. metabolic activity) and longer time scales (e.g. evolutionary selection), along with their complex spatial organization. Because of their complexity and their ability to innovate with respect to their environment, living systems are considered to be open-ended. Historically, it has been difficult to model open-ended evolution and innovation. As a result, our understanding of the exact mechanisms that distinguish open-ended living systems from non-living ones is limited. One of the biggest barriers is understanding how multiple, complex parts within a single system interact and contribute to the complex, emergent behavior of the system as a whole. How do interactions between parts of a system lead to more complex behavior of the system as a whole? This paper presents two interacting cellular automata (CA) as an abstract model to address the effects of complex interactions between two individual entities embedded within a larger system. Unlike elementary CA, each CA changes their update rules as a function of the system’s state as a whole. The resulting behavior of the two-CA system suggests that complex interaction functions between the two CA have little to no effect on the complexity of each individual CA behavior and structure. However, having an interaction function that is random results in open-ended evolution regardless of the specific type of state-dependency.

\end{abstract}
\maketitle

\section{Introduction}

Biological systems are notorious for their complex physical structures and complex behaviors over short and long time scales. Biological systems also tend to partition themselves into smaller subsystems that interact with other subsystems. For example, the human gut microbiome consists of several bacterial and viral communities. Viruses of bacteria (phages) are being recognized as important components of the human microbiome due to their interaction effects with bacterial communities\cite{10.3389/fendo.2019.00784}. They modulate bacterial communities by killing bacteria and driving metabolic activity. However, little is known about the specific roles played by phages in human systems, particularly how exactly they interact with bacterial communities, how that plays a role in the human gut health overall, and how those effects in turn impact the phage communities. In general, the exact implications on system structure and dynamics of interacting subsystems, particularly when it comes to measurable biological complexity and behavior, is not well-understood\cite{WolfE8678, 10.1093/icb/icv033}.

Biology is also known to evolve open-endedly, meaning it continuously innovates over time (in several different ways), maintains a certain amount of complexity, and never repeats itself exactly\cite{bedau2000open, banzhaf2016defining, ruiz2008enabling, taylor1999artificial, taylor2016open, 10.1162/artl_a_00291, dolsoninno}. However, there is no wide agreement on a precise definition of open-endedness in the literature\cite{taylor2016open, 10.1162/artl_a_00291}. The relationship between a system's capacity for open-endedness and complexity also remains imprecise, even within computational models.

It is widely known that interacting components of a system can lead to emergent behavior of the system as a whole\cite{Chalmers2006-CHASAW, Bedau2008EmergenceC}. Bedau\cite{https://doi.org/10.1111/0029-4624.31.s11.17} recognizes emergence in two main forms, weak and strong. In weak emergence, the behavior of individual entities sums exactly to completely describe the behavior of a group. This emergence can be derived by simulating internal dynamics and known external conditions. Strong emergence, on the other hand, is more difficult to understand because it cannot be simulated from internal dynamics and known external conditions. Both forms are closely related to the concept of innovation used to define open-endedness. Here, something entirely new must emerge from underlying dynamics, whether it be a new pattern/structure or an entirely new set of rules altogether\cite{10.1162/isal_a_00382}. Understanding the mechanisms of open-endedness, particularly how open-ended systems innovate in several different ways, might be useful for understanding mechanisms that drive emergent phenomena.

Biological systems have natural partitions that define subsystems, such as individual organisms, individual cells, different cell types, and different species. From a bottom-up approach, it is assumed that a system's behavior is entirely determined by the underlying laws of the parts that compose it\cite{ellis2011top, walker2012evolutionary}. A top-down approach suggests just the opposite: That the behavior of individual parts in a system is determined by the behavior of the system as a whole\cite{ellis2011top, walker2012evolutionary, topdown}. Researchers use either (or both) approaches to explain how entities within a system affect system dynamics over time\cite{me2017, walker2012evolutionary}. As an example, human health is impacted by interactions that occur between phages and microbes. In turn, an individual's overall health impacts their behavior, including diet, which impacts the communities of phages and microbes.

\subsection{Prior Work}

The main goal of this paper is to explore the relationship between interaction functions between subsystems, complexity, and open-endedness within a simple computational model. One form of open-ended evolution (defined in Section~\ref{sec:oee}) has been demonstrated within a system of interacting elementary cellular automata\cite{me2017}. In that model, one cellular automata (CA) evolves according to a fixed rule, but a second evolves according to a rule that can change at each time step. The rule to evolve the next state is determined by the current state of both spatially separate CA under an arbitrary interaction function $f$. As a result, the CA that changes its rule can evolve open-endedly in the sense that it takes longer than expected to repeat its spatial pattern and also that the pattern is innovative with respect to all possible patterns generated by a fixed rule ($r \in R$ where $R$ is the set of 256 ECA rules).

In addition, results from the previous model show that this form of open-endedness is not scalable with system size if the updating rule changes randomly, without considering the current state of both CA. No open-endedness was observed if the rule was only determined by a single CA, suggesting that open-endedness may depend on state-dependent dynamics of the system as a whole. Finally, assuming the transitions between states can be represented on a directed graph, the topology of this interacting system allows walks along states and edges that are innovative and open-ended. This is because a wider variety of states and edges are more accessible at each time step\cite{e19090461}.

This paper extends these results by allowing both CA to change their updating rule at each time step. In this experiment, I explore a few different interaction functions and measure the complexity of the resulting spatial CA patterns along with the trajectory of states over time for each CA. Due to the algorithmic nature of this model, approximations to algorithmic complexity are used in place of entropy-based measures of complexity throughout this analysis\cite{hectormain}.

\section{\label{sec:oee}Apparent Open-Ended Evolution}

Even though biological evolution is widely accepted as being open-ended\cite{packard1, taylor2016open}, there is no scientific consensus on an exact, quantifiable definition of open-endedness\cite{taylor2016open, 10.1162/artl_a_00291}. But for bounded, discrete dynamical models with synchronous update rules, the open-ended evolution (OEE) of states over time is defined in terms of innovation and unbounded evolution\cite{me2017}. These are defined in Adams, et al 2017\cite{me2017} and are also defined below as abbreviated versions.

The Poincar\'e recurrence time $t_P$ of a finite, deterministic, bounded, and dynamical model provides a time constraint on when it will repeat exactly. For 1-dimensional ECA, $t_P = 2^w$, where $w$ is the number of cells in a single CA state. This is because an ECA can, at most, express every possible state in its evolutionary state trajectory, and because the update rule $r$ is fixed, visiting the same state more than once would cause the state trajectory to repeat itself exactly.

{\it Unbounded evolution}\cite{me2017} is the ability for a single CA model $c$ embedded within a larger system $u$ of two interacting CA ($c_1$ and $c_2$) to defy the Poincar\'e recurrence time by repeating its patterns of expressed states $s(t_1),s(t_2),s(t_3) \ldots s(t_r)$ in a time greater than $t_P$. Because $c_1$ and $c_2$ within $u$ are \textit{not} isolated and do \textit{not} evolve under a single, unchanging update rule like ECA, the $t_P$ that $c$ needs to ``beat'' is determined by the $t_P$ of an ECA with the same size $w$ as $c$.

\bigskip 
\begin{mydef} \label{Def:UE}
{\bf Unbounded evolution}: A finite, deterministic, and bounded dynamical system $u$, which can be decomposed into subsystems $c_1$ and $c_2$ that interact according to a function $f$, exhibits unbounded evolution if there exists a recurrence time $t_r$ in $c_1$ or $c_2$ such that $s_{f}(t)=s(t_1),s(t_2),s(t_3) \ldots s(t_r)$ is non-repeating for $t_r>t_P$, where $t_P$ is the Poincar\'e recurrence time for an equivalent isolated (non-interacting) system $c$.
\end{mydef}
\bigskip 

Here, $f$ is the interaction function between $c_1$ and $c_2$, defined in Section~\ref{sec:model}. Because the state evolution (trajectory) of $c$ is compared to counterfactual state trajectories of ECA of the same size, this implies that ECA are inherently incapable of unbounded evolution. \textit{Innovation} is defined as\cite{me2017}:

\bigskip 
\begin{mydef} \label{Def:INN}
{\bf Innovation}: A finite, deterministic, and bounded dynamical system $u$, which can be decomposed into subsystems $c_1$ and $c_2$ that interact according to a function $f$, exhibits innovation if there exists a state trajectory $s_{f}(t)=s(t_1),s(t_2),s(t_3) \ldots s(t_r)$ that is not contained in the set of all possible state trajectories $\{ s_I \}$ for an equivalent isolated (non-interacting) system $c$. 
\end{mydef}
\bigskip 

That is, a subsystem CA $c$ exhibits innovation by Definition~\ref{Def:INN} if its state trajectory {\it cannot} be produced by an ECA of the same size. Both Definitions~\ref{Def:UE} and~\ref{Def:INN} reflect the intuitive notions of ``on-going production of novelty'' and ``unbounded evolution''\cite{banzhaf2016defining} but do not necessarily mean the complexity of individual states increases with time. Furthermore, OEE is apparent on the scale of a single CA embedded within a larger system. This is in agreement with our intuition of OEE within biology--- the evolution of life as a whole appears to evolve open-endedly, but it is embedded within a larger system that is not necessarily open-ended.

\section{\label{sec:model}Model}

\begin{figure}[h!]
\centering
\includegraphics[width=0.8\linewidth]{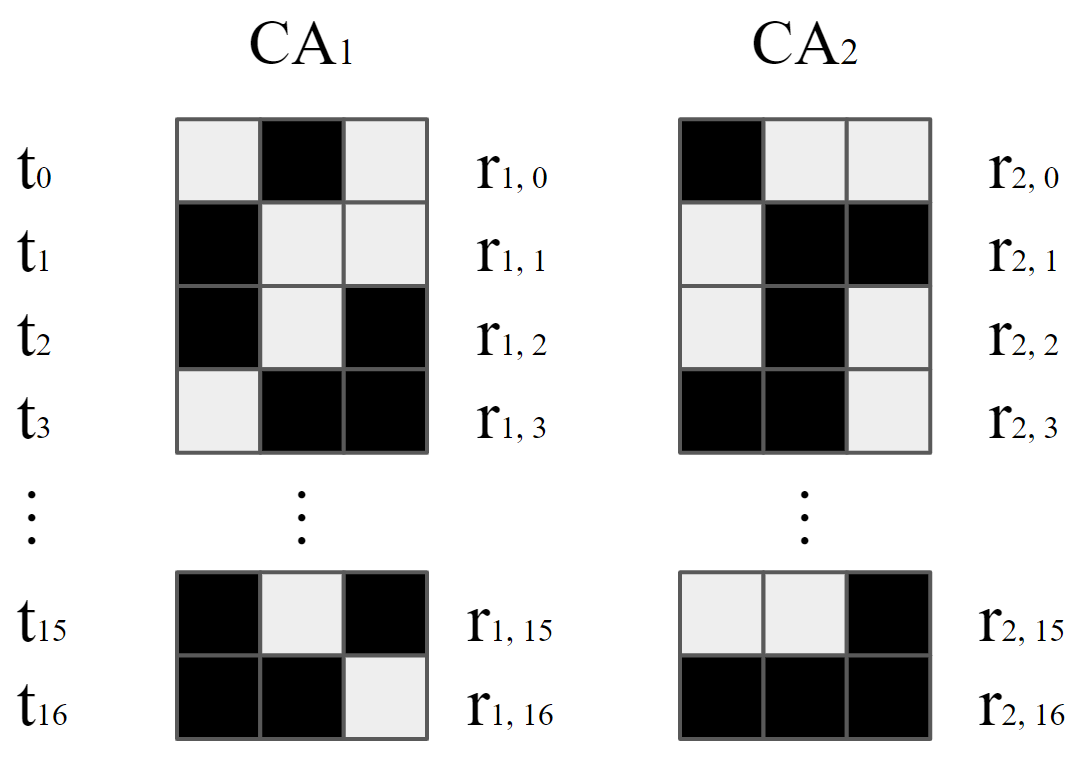}
\caption{Each of these CA evolve over time (downwards) and change their update rule (one of the 256 ECA rules) according to some function $f_{type}$ as described in the text. Because results are exhaustive, only CA with $w=3$ are considered.}
\label{fig:ca}
\end{figure}

The model system explored here is composed of two finite, deterministic, and spatially bounded interacting CA with fixed widths $w$ and periodic boundary conditions. Each CA starts exactly like an ECA with a fixed update rule (one of the 256 ECA rules), but after each time step, each CA changes its rule that is used to determine that CA's next state $s$. Both CA use one of the following types of functions $f_{\text{type}, t}$ at time $t$ to change their rule $r$, which will then determine the next state $s_{t+1}$: 

\begin{enumerate}
  \item $r_{x, t} = f_{\text{this state}, t}(s_{x, t})$: $r_{x, t}$ is determined only by the current state $s_{x, t}$ of that CA.
  \item $r_{x, t} = f_{\text{other state}, t}(s_{y, t})$: $r_{x, t}$ is determined only by the current state $s_{y, t}$ of the opposing CA.
  \item $r_{1, t} = r_{2, t} = f_{\text{both states}, t}(s_{1, t}, s_{2, t})$: Both $r_{1, t}$ and $r_{2, t}$ are determined by the current states $s_t$ of both CA.
  \item $r_{x, t} = f_{\text{mixed}, t}(\text{random choice}((s_{1, t}, s_{2, t}) \lor s_{x, t} \lor s_{y, t}))$: $r_{x, t}$ depends on a random choice of both CA states, the state of that CA, or the state of the opposing CA.
\end{enumerate}

Because all possible initial states for a given $w$ and initial rules are explored for both CA, only CA with $w=3$ were explored. For each of all possible combinations of $s_{0, w}$, $r_0$, state trajectories were recorded for $2*2^w$ time steps. An illustration of this model is shown in Figure~\ref{fig:ca}.

For each interaction function type $f_{\text{type}}$, the exact mappings $f_{\text{type}, i}$ between states $s$ and rules $r$ was generated randomly. 5000 random mappings were created for each $f_{\text{type}}$, and only six of these 5000 mappings were used. The six mappings were chosen based on their relative approximate complexity values (described in~\ref{sec:methods})--- three mappings with relatively high complexity and three mappings with relatively low complexity. Because the interaction function for $f_{\text{mixed}}$ depends on a random choice made at each time step $t$, the complexity could not be measured for $f_{\text{mixed}, i}$ mappings, since a static mapping does not exist. However, the exact random choice in mappings were based on the other three interaction types. For each of the interaction function types, the six individual mappings are denoted as $i \in {0, 1, 2, 3, 4, 5}$.

\section{\label{sec:methods}Methods}

Both CA state trajectories were checked for apparent OEE according to Definitions~\ref{Def:UE} and~\ref{Def:INN}. Algorithmic complexity (Kolmogorov complexity) cannot be computed exactly due to the Halting Problem, but can be approximated using the Block Decomposition Method (BDM)\cite{hectormain}. This is an upper-bound approximation of the algorithmic complexity, which, in short, measures the size of the smallest computer program that can produce the string of symbols being measured\cite{hectormain}.

The BDM can be used to approximate the algorithmic complexity of 1-dimensional or 2-dimensional objects. By representing each interaction function mapping $f_{\text{type}, i}$ as an adjacency matrix, it is possible to approximate the complexity for any non-changing $f_{\text{type}, i}$\cite{hector2}. But since each interaction function mapping $f_{\text{type}, i}$ was generated randomly, this would affect the expected range of BDM values for any $f_{\text{type}, i}$. BDM is largely known for quantifying the randomness of mechanisms capable of producing an object. This was mitigated by selecting mappings with high and low BDM values \textit{relative} to a batch of 5000 randomly-generated mappings.

The CA state trajectory of both CA can be represented in two ways for the purposes of measuring the BDM. The first is to measure the BDM of each state in the state trajectory. Then the \textit{mean} of the BDM values for each individual state $s$ are calculated per CA state trajectory. The second is by enumerating all possible states for a CA of size $w$ and measuring the BDM of the sequence of enumerated states. For $w=3$, there are $2^w=8$ possible states, making it computationally tractable to measure the BDM for the entire state trajectory. For computational tractability reasons, the Python 3 package \textit{pybdm} (https://pybdm-docs.readthedocs.io/en/latest/) does not support sequences with an alphabet size over 9, thus making it possible to use \textit{pybdm} to calculate the BDM for each of these measurements.

\newpage
\section{\label{sec:results}Results}

\begin{figure}[ht!]
\centering
\includegraphics[width=1.0\linewidth]{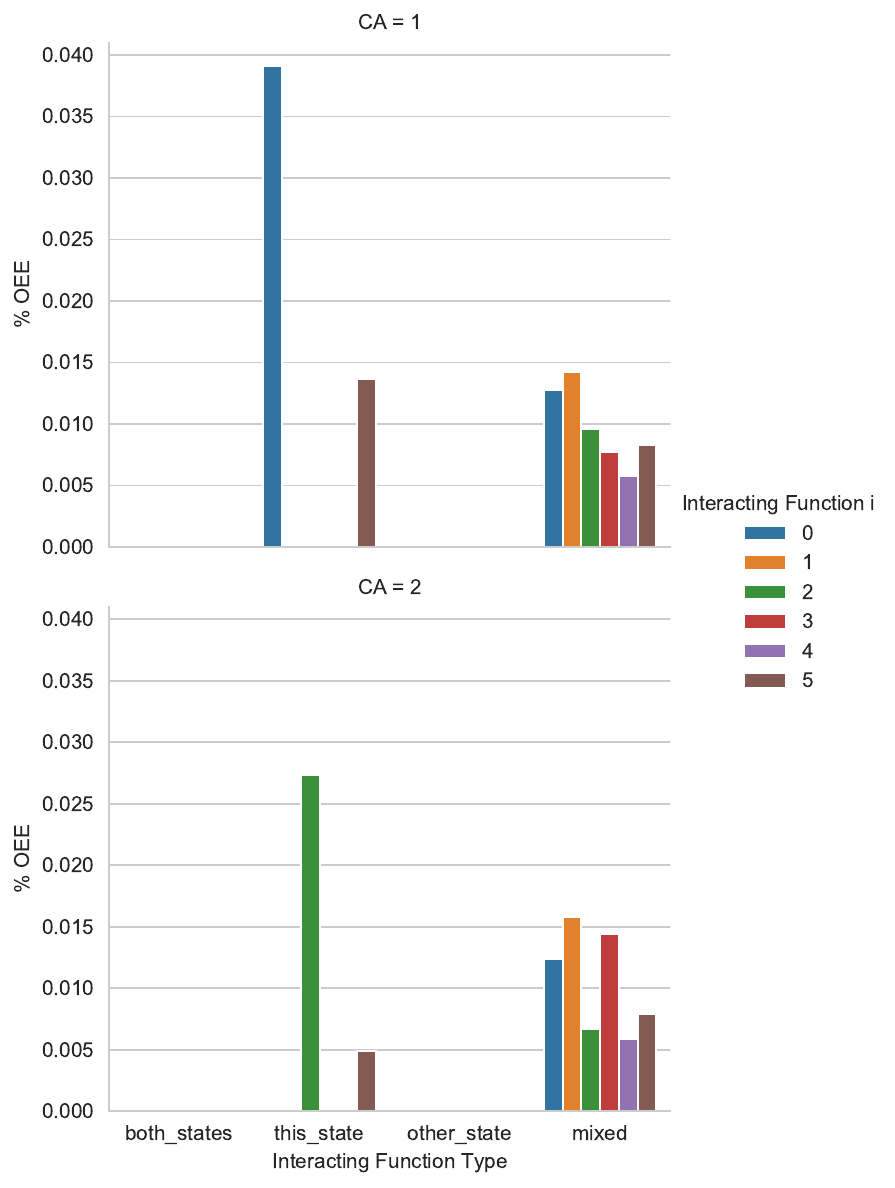}
\caption{Percent of all OEE state trajectories for all interaction function types. Results are only shown for CA 1 trajectories; results for CA 2 are similar because the system is symmetric.}
\label{fig:oee}
\end{figure}

Figure~\ref{fig:oee} shows the number of open-ended CA state trajectories (\% OEE) for both CA as a function of the different interaction function types $f_{\text{type}}$, as defined by Definitions~\ref{Def:UE} and~\ref{Def:INN}. There were no OEE state trajectories for $f_{\text{both states}}$, since the $t_r$ of each CA depends on the states and rules from both CA simultaneously. Then, by definition, there are no OEE state trajectories for $f_{\text{both states}}$ since the $t_r$ is completely determined by the dynamics of the whole system. There were also no OEE state trajectories $f_{\text{other state}}$. This is consistent with prior results\cite{me2017}, which suggest that the open-ended evolution of a subsystem is largely dependent on state-dependent dynamics-- it \textit{must} use its own state in $f_{\text{type}}$ to produce open-ended behavior.

\begin{figure}[ht!]
\centering
\includegraphics[width=1.0\linewidth]{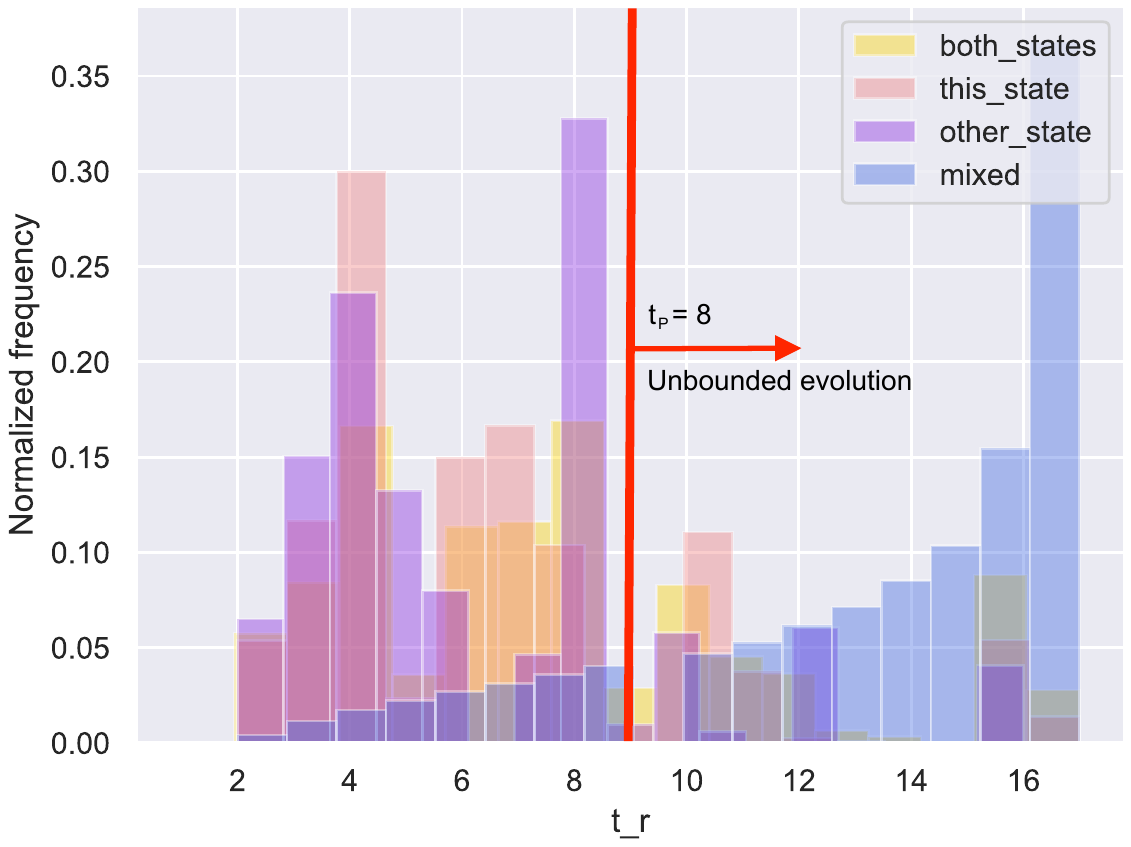}
\caption{Distributions of $t_r$ values for each interaction function type. CA state trajectories with $t_r>t_P$ exhibit unbounded evolution, as defined in Definition~\ref{Def:UE}.}
\label{fig:tr}
\end{figure}

The distributions of recurrence times $t_r$ are shown in Figure~\ref{fig:tr}. If the $t_r > 2*2^w$, then $t_r$ was denoted as $t_r = 2*2^w + 1$, for computational simplicity. Within each $f_{\text{type}}$, the individual mappings showed little difference in the distribution of $t_r$. The interaction function $f_{\text{mixed}}$ shows an exponential-like distribution.

\begin{figure}[ht!]
\centering
\includegraphics[width=1.0\linewidth]{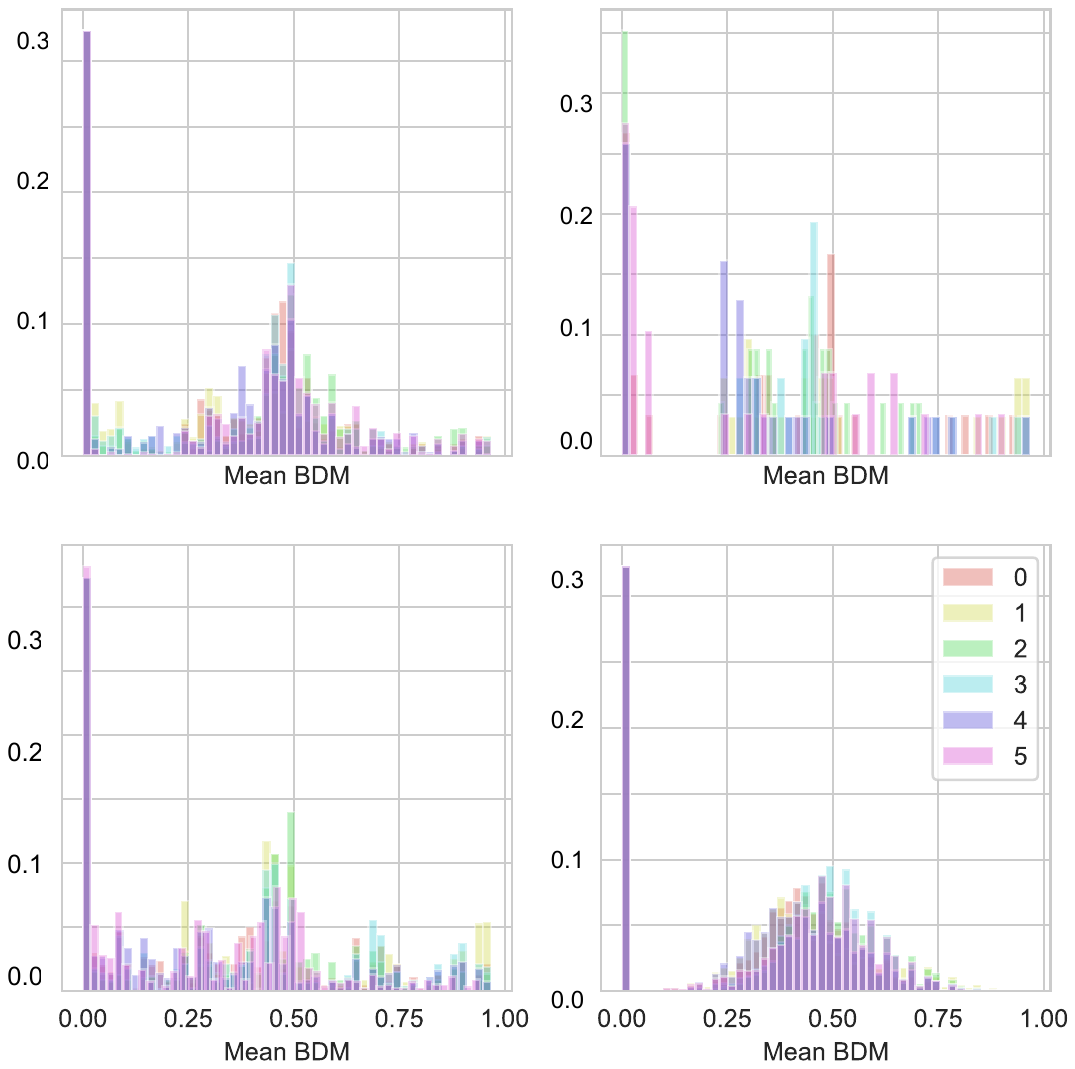}
\caption{Distributions of mean BDM values of individual states in a CA state trajectory for each interaction function type (both states, this state, other state, mixed, from upper left to lower right, respectively).}
\label{fig:bdm_dists_mean_panel_1}
\end{figure}

\begin{figure}[ht!]
\centering
\includegraphics[width=1.0\linewidth]{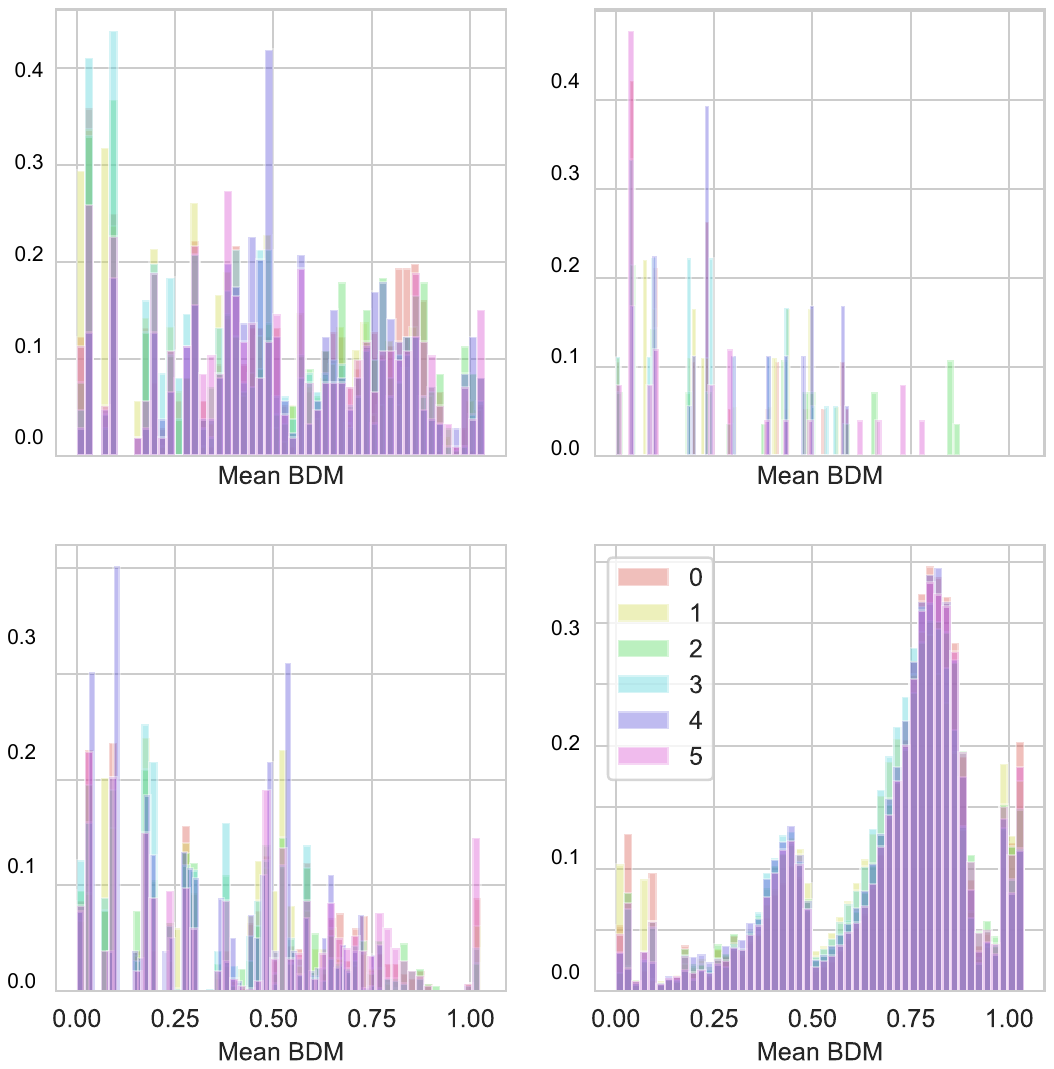}
\caption{Distributions of mean BDM values of enumerated state trajectories in a CA for each interaction function type (both states, this state, other state, mixed, from upper left to lower right, respectively).}
\label{fig:bdm_dists_traj_panel_1}
\end{figure}

The distributions of the mean BDM of a state within a CA state trajectory is shown in Figure~\ref{fig:bdm_dists_mean_panel_1} for each $f_{\text{type}}$. Similarly, the distribution of the BDM for each enumerated state trajectory is show in Figure~\ref{fig:bdm_dists_traj_panel_1}. These results are for CA 1, and results for CA 2 are very similar because the system is symmetric (not shown).

\begin{figure}[ht!]
\centering
\includegraphics[width=1.1\linewidth]{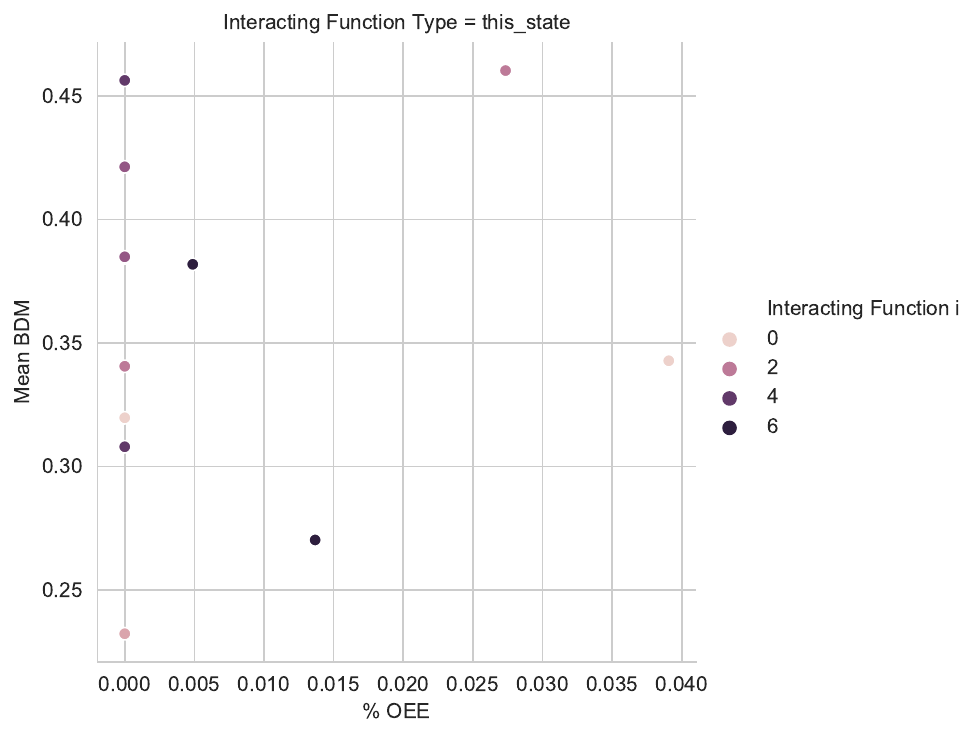}
\caption{\% OEE state trajectories vs. average mean BDM values of individual states in a CA state trajectory for all six $f_{\text{this state}, i}$ mappings.}
\label{fig:bdm_dists_mean_vs_oee}
\end{figure}

\begin{figure}[ht!]
\centering
\includegraphics[width=1.0\linewidth]{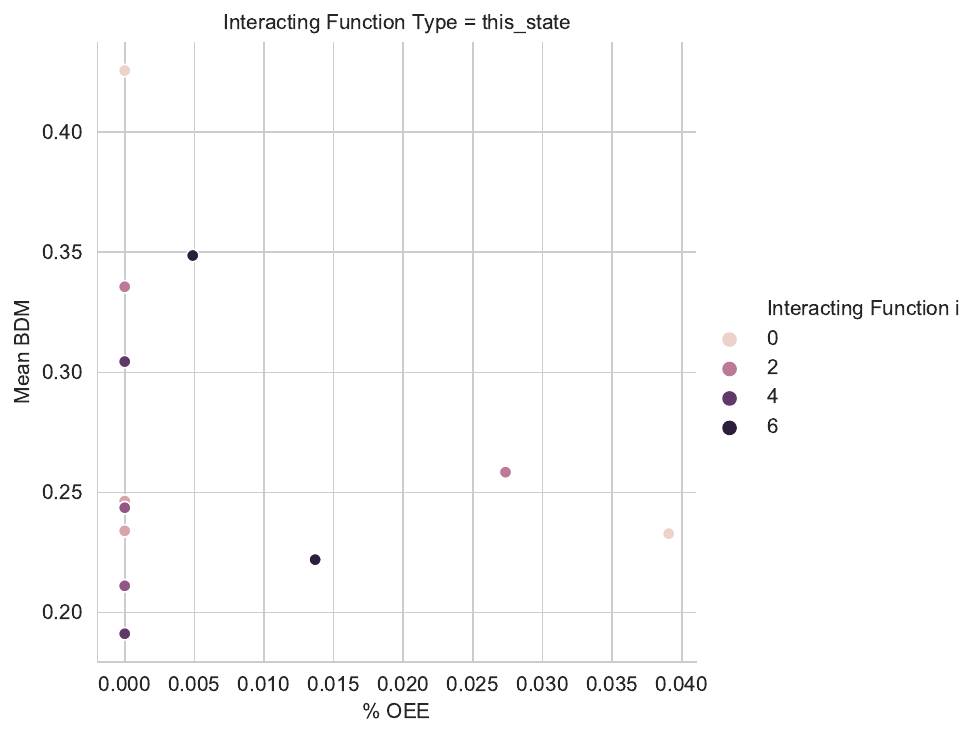}
\caption{\% OEE state trajectories vs. average BDM values of enumerated CA state trajectories for all six $f_{\text{this state}, i}$ mappings.}
\label{fig:bdm_dists_traj_vs_oee}
\end{figure}

For $f_{\text{this state}}$, the relationship between \% OEE and mean BDM are shown in Figures~\ref{fig:bdm_dists_mean_vs_oee} and~\ref{fig:bdm_dists_traj_vs_oee} for the mean BDM of states within a trajectory and the enumerated state trajectory, respectively. The average mean BDM value is separated by the six mappings $f_{\text{this state}, i}$. The other interaction function types are not shown because either they were not able to produce OEE state trajectories, or because the BDM for the mapping could not be measured (for $f_{\text{mixed}, i}$ mappings). Results for CA 2 are similar and are not shown.

\begin{figure}[ht!]
\centering
\includegraphics[width=0.9\linewidth]{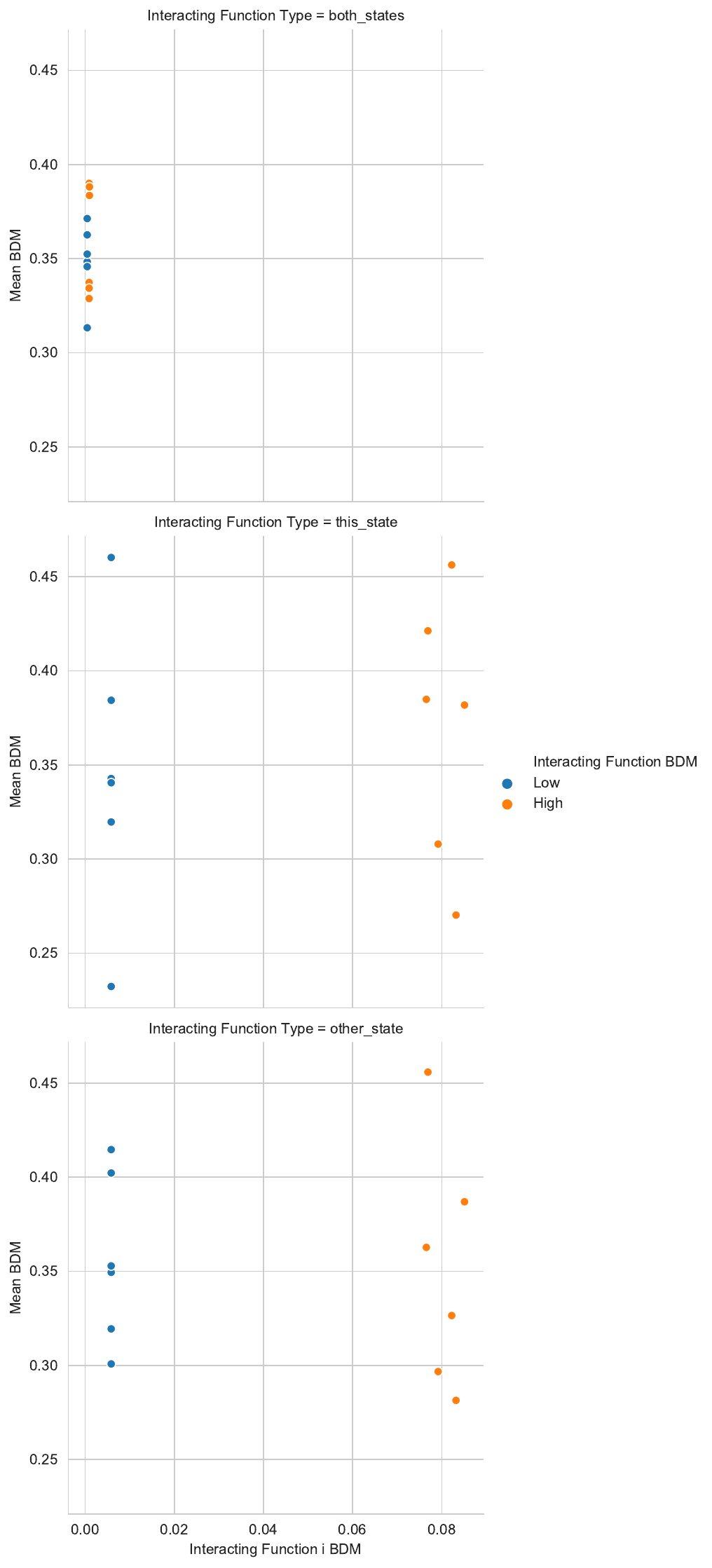}
\caption{BDM complexity values for each $f_{\text{type}, i}$ mapping vs. average mean BDM values of individual states in a CA state trajectory.}
\label{fig:bdm_dists_mean_vs_bdm}
\end{figure}

\begin{figure}[ht!]
\centering
\includegraphics[width=0.9\linewidth]{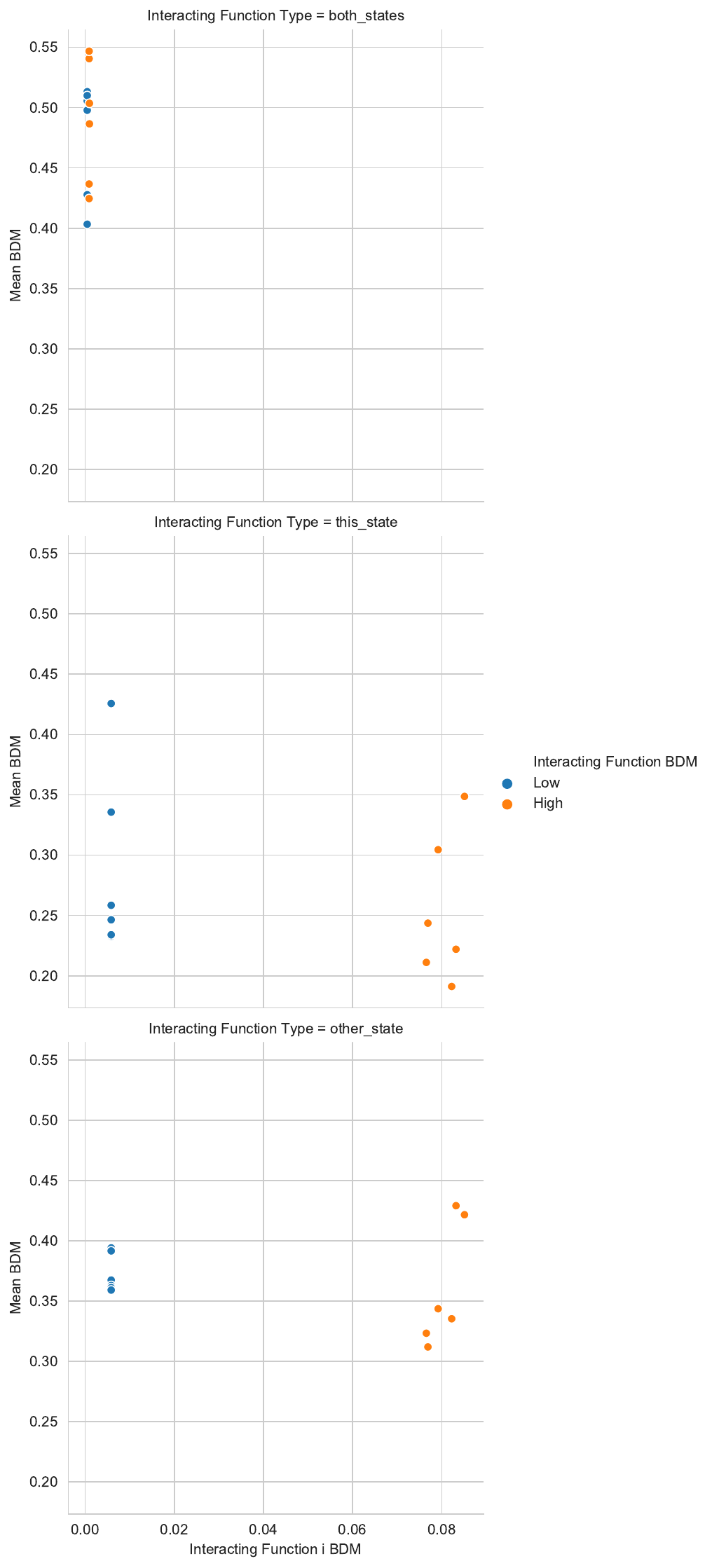}
\caption{BDM complexity values for each $f_{\text{type}, i}$ mapping vs. average BDM values of enumerated CA state trajectories.}
\label{fig:bdm_dists_traj_vs_bdm}
\end{figure}

Finally, Figures~\ref{fig:bdm_dists_mean_vs_bdm} and~\ref{fig:bdm_dists_traj_vs_bdm} show the $f_{\text{type}, i}$ mapping BDMs vs. the mean BDM of states within a trajectory and the enumerated state trajectory, respectively, for the three $f_{\text{type}}$ with measurable BDMs. Each panel is plotted with the same x and y range for comparison.

\newpage
\section{\label{sec:discussion}Discussion}

These results suggest that the complexity of an interaction function (mappings between system states and CA rules for the next time step) likely has \textit{little to no effect} on the complexity of the behavior of an individual CA embedded in a larger system. However, Figure~\ref{fig:oee}, Figure~\ref{fig:tr}, and Figure~\ref{fig:bdm_dists_traj_panel_1} strongly suggest that an interaction function between two CA that randomly changes its state-dependencies results in open-ended evolution, complex states, \textit{and} complex state dynamics, regardless of the static mapping between states and rules it chooses from. This suggests the results for $f_{\text{mixed}}$ are robust against the exact mapping between states and rules, and the resulting complex behavior within a CA results from the random dependencies of different parts of the entire system.

For real biological systems, such as a human gut microbiome, these results suggest that the complex evolutionary behavior of individual communities may not be a consequence of a fixed, static relationship between entities, but rather of random events from multiple parts of an entire system. These results could guide data-driven empirical analyses on how communities, individuals, or other entities interact to form complex behavior in real biological systems.

\newpage
 
\newpage
\bibliography{bib}

\begin{thebibliography}{22}%
\makeatletter
\providecommand \@ifxundefined [1]{%
 \@ifx{#1\undefined}
}%
\providecommand \@ifnum [1]{%
 \ifnum #1\expandafter \@firstoftwo
 \else \expandafter \@secondoftwo
 \fi
}%
\providecommand \@ifx [1]{%
 \ifx #1\expandafter \@firstoftwo
 \else \expandafter \@secondoftwo
 \fi
}%
\providecommand \natexlab [1]{#1}%
\providecommand \enquote  [1]{``#1''}%
\providecommand \bibnamefont  [1]{#1}%
\providecommand \bibfnamefont [1]{#1}%
\providecommand \citenamefont [1]{#1}%
\providecommand \href@noop [0]{\@secondoftwo}%
\providecommand \href [0]{\begingroup \@sanitize@url \@href}%
\providecommand \@href[1]{\@@startlink{#1}\@@href}%
\providecommand \@@href[1]{\endgroup#1\@@endlink}%
\providecommand \@sanitize@url [0]{\catcode `\\12\catcode `\$12\catcode
  `\&12\catcode `\#12\catcode `\^12\catcode `\_12\catcode `\%12\relax}%
\providecommand \@@startlink[1]{}%
\providecommand \@@endlink[0]{}%
\providecommand \url  [0]{\begingroup\@sanitize@url \@url }%
\providecommand \@url [1]{\endgroup\@href {#1}{\urlprefix }}%
\providecommand \urlprefix  [0]{URL }%
\providecommand \Eprint [0]{\href }%
\providecommand \doibase [0]{http://dx.doi.org/}%
\providecommand \selectlanguage [0]{\@gobble}%
\providecommand \bibinfo  [0]{\@secondoftwo}%
\providecommand \bibfield  [0]{\@secondoftwo}%
\providecommand \translation [1]{[#1]}%
\providecommand \BibitemOpen [0]{}%
\providecommand \bibitemStop [0]{}%
\providecommand \bibitemNoStop [0]{.\EOS\space}%
\providecommand \EOS [0]{\spacefactor3000\relax}%
\providecommand \BibitemShut  [1]{\csname bibitem#1\endcsname}%
\let\auto@bib@innerbib\@empty
\bibitem [{\citenamefont {Sutton}\ and\ \citenamefont
  {Hill}(2019)}]{10.3389/fendo.2019.00784}%
  \BibitemOpen
  \bibfield  {author} {\bibinfo {author} {\bibfnamefont {T.~D.~S.}\
  \bibnamefont {Sutton}}\ and\ \bibinfo {author} {\bibfnamefont
  {C.}~\bibnamefont {Hill}},\ }\href {\doibase 10.3389/fendo.2019.00784}
  {\bibfield  {journal} {\bibinfo  {journal} {Frontiers in Endocrinology}\
  }\textbf {\bibinfo {volume} {10}},\ \bibinfo {pages} {784} (\bibinfo {year}
  {2019})}\BibitemShut {NoStop}%
\bibitem [{\citenamefont {Wolf}\ \emph {et~al.}(2018)\citenamefont {Wolf},
  \citenamefont {Katsnelson},\ and\ \citenamefont {Koonin}}]{WolfE8678}%
  \BibitemOpen
  \bibfield  {author} {\bibinfo {author} {\bibfnamefont {Y.~I.}\ \bibnamefont
  {Wolf}}, \bibinfo {author} {\bibfnamefont {M.~I.}\ \bibnamefont
  {Katsnelson}}, \ and\ \bibinfo {author} {\bibfnamefont {E.~V.}\ \bibnamefont
  {Koonin}},\ }\href {\doibase 10.1073/pnas.1807890115} {\bibfield  {journal}
  {\bibinfo  {journal} {Proceedings of the National Academy of Sciences}\
  }\textbf {\bibinfo {volume} {115}},\ \bibinfo {pages} {E8678} (\bibinfo
  {year} {2018})},\ \Eprint
  {http://arxiv.org/abs/https://www.pnas.org/content/115/37/E8678.full.pdf}
  {https://www.pnas.org/content/115/37/E8678.full.pdf} \BibitemShut {NoStop}%
\bibitem [{\citenamefont {Kane}\ and\ \citenamefont
  {Higham}(2015)}]{10.1093/icb/icv033}%
  \BibitemOpen
  \bibfield  {author} {\bibinfo {author} {\bibfnamefont {E.~A.}\ \bibnamefont
  {Kane}}\ and\ \bibinfo {author} {\bibfnamefont {T.~E.}\ \bibnamefont
  {Higham}},\ }\href {\doibase 10.1093/icb/icv033} {\bibfield  {journal}
  {\bibinfo  {journal} {Integrative and Comparative Biology}\ }\textbf
  {\bibinfo {volume} {55}},\ \bibinfo {pages} {146} (\bibinfo {year} {2015})},\
  \Eprint
  {http://arxiv.org/abs/https://academic.oup.com/icb/article-pdf/55/1/146/17056208/icv033.pdf}
  {https://academic.oup.com/icb/article-pdf/55/1/146/17056208/icv033.pdf}
  \BibitemShut {NoStop}%
\bibitem [{\citenamefont {Bedau}\ \emph {et~al.}(2000)\citenamefont {Bedau},
  \citenamefont {McCaskill}, \citenamefont {Packard}, \citenamefont
  {Rasmussen}, \citenamefont {Adami}, \citenamefont {Green}, \citenamefont
  {Ikegami}, \citenamefont {Kaneko},\ and\ \citenamefont
  {Ray}}]{bedau2000open}%
  \BibitemOpen
  \bibfield  {author} {\bibinfo {author} {\bibfnamefont {M.~A.}\ \bibnamefont
  {Bedau}}, \bibinfo {author} {\bibfnamefont {J.~S.}\ \bibnamefont
  {McCaskill}}, \bibinfo {author} {\bibfnamefont {N.~H.}\ \bibnamefont
  {Packard}}, \bibinfo {author} {\bibfnamefont {S.}~\bibnamefont {Rasmussen}},
  \bibinfo {author} {\bibfnamefont {C.}~\bibnamefont {Adami}}, \bibinfo
  {author} {\bibfnamefont {D.~G.}\ \bibnamefont {Green}}, \bibinfo {author}
  {\bibfnamefont {T.}~\bibnamefont {Ikegami}}, \bibinfo {author} {\bibfnamefont
  {K.}~\bibnamefont {Kaneko}}, \ and\ \bibinfo {author} {\bibfnamefont {T.~S.}\
  \bibnamefont {Ray}},\ }\href@noop {} {\bibfield  {journal} {\bibinfo
  {journal} {Artificial life}\ }\textbf {\bibinfo {volume} {6}},\ \bibinfo
  {pages} {363} (\bibinfo {year} {2000})}\BibitemShut {NoStop}%
\bibitem [{\citenamefont {Banzhaf}\ \emph {et~al.}(2016)\citenamefont
  {Banzhaf}, \citenamefont {Baumgaertner}, \citenamefont {Beslon},
  \citenamefont {Doursat}, \citenamefont {Foster}, \citenamefont {McMullin},
  \citenamefont {De~Melo}, \citenamefont {Miconi}, \citenamefont {Spector},
  \citenamefont {Stepney} \emph {et~al.}}]{banzhaf2016defining}%
  \BibitemOpen
  \bibfield  {author} {\bibinfo {author} {\bibfnamefont {W.}~\bibnamefont
  {Banzhaf}}, \bibinfo {author} {\bibfnamefont {B.}~\bibnamefont
  {Baumgaertner}}, \bibinfo {author} {\bibfnamefont {G.}~\bibnamefont
  {Beslon}}, \bibinfo {author} {\bibfnamefont {R.}~\bibnamefont {Doursat}},
  \bibinfo {author} {\bibfnamefont {J.~A.}\ \bibnamefont {Foster}}, \bibinfo
  {author} {\bibfnamefont {B.}~\bibnamefont {McMullin}}, \bibinfo {author}
  {\bibfnamefont {V.~V.}\ \bibnamefont {De~Melo}}, \bibinfo {author}
  {\bibfnamefont {T.}~\bibnamefont {Miconi}}, \bibinfo {author} {\bibfnamefont
  {L.}~\bibnamefont {Spector}}, \bibinfo {author} {\bibfnamefont
  {S.}~\bibnamefont {Stepney}},  \emph {et~al.},\ }\href@noop {} {\bibfield
  {journal} {\bibinfo  {journal} {Theory in Biosciences}\ ,\ \bibinfo {pages}
  {1}} (\bibinfo {year} {2016})}\BibitemShut {NoStop}%
\bibitem [{\citenamefont {Ruiz-Mirazo}\ \emph {et~al.}(2008)\citenamefont
  {Ruiz-Mirazo}, \citenamefont {Umerez},\ and\ \citenamefont
  {Moreno}}]{ruiz2008enabling}%
  \BibitemOpen
  \bibfield  {author} {\bibinfo {author} {\bibfnamefont {K.}~\bibnamefont
  {Ruiz-Mirazo}}, \bibinfo {author} {\bibfnamefont {J.}~\bibnamefont {Umerez}},
  \ and\ \bibinfo {author} {\bibfnamefont {A.}~\bibnamefont {Moreno}},\
  }\href@noop {} {\bibfield  {journal} {\bibinfo  {journal} {Biology \&
  Philosophy}\ }\textbf {\bibinfo {volume} {23}},\ \bibinfo {pages} {67}
  (\bibinfo {year} {2008})}\BibitemShut {NoStop}%
\bibitem [{\citenamefont {Taylor}(1999)}]{taylor1999artificial}%
  \BibitemOpen
  \bibfield  {author} {\bibinfo {author} {\bibfnamefont {T.~J.}\ \bibnamefont
  {Taylor}},\ }\emph {\bibinfo {title} {From artificial evolution to artificial
  life}},\ \href@noop {} {Ph.D. thesis},\ \bibinfo  {school} {University of
  Edinburgh. College of Science and Engineering. School of Informatics.}
  (\bibinfo {year} {1999})\BibitemShut {NoStop}%
\bibitem [{\citenamefont {Taylor}\ \emph {et~al.}(2016)\citenamefont {Taylor},
  \citenamefont {Bedau}, \citenamefont {Channon}, \citenamefont {Ackley},
  \citenamefont {Banzhaf}, \citenamefont {Beslon}, \citenamefont {Dolson},
  \citenamefont {Froese}, \citenamefont {Hickinbotham}, \citenamefont {Ikegami}
  \emph {et~al.}}]{taylor2016open}%
  \BibitemOpen
  \bibfield  {author} {\bibinfo {author} {\bibfnamefont {T.}~\bibnamefont
  {Taylor}}, \bibinfo {author} {\bibfnamefont {M.}~\bibnamefont {Bedau}},
  \bibinfo {author} {\bibfnamefont {A.}~\bibnamefont {Channon}}, \bibinfo
  {author} {\bibfnamefont {D.}~\bibnamefont {Ackley}}, \bibinfo {author}
  {\bibfnamefont {W.}~\bibnamefont {Banzhaf}}, \bibinfo {author} {\bibfnamefont
  {G.}~\bibnamefont {Beslon}}, \bibinfo {author} {\bibfnamefont
  {E.}~\bibnamefont {Dolson}}, \bibinfo {author} {\bibfnamefont
  {T.}~\bibnamefont {Froese}}, \bibinfo {author} {\bibfnamefont
  {S.}~\bibnamefont {Hickinbotham}}, \bibinfo {author} {\bibfnamefont
  {T.}~\bibnamefont {Ikegami}},  \emph {et~al.},\ }\href@noop {} {\bibfield
  {journal} {\bibinfo  {journal} {Artificial Life}\ } (\bibinfo {year}
  {2016})}\BibitemShut {NoStop}%
\bibitem [{\citenamefont {Packard}\ \emph {et~al.}(2019)\citenamefont
  {Packard}, \citenamefont {Bedau}, \citenamefont {Channon}, \citenamefont
  {Ikegami}, \citenamefont {Rasmussen}, \citenamefont {Stanley},\ and\
  \citenamefont {Taylor}}]{10.1162/artl_a_00291}%
  \BibitemOpen
  \bibfield  {author} {\bibinfo {author} {\bibfnamefont {N.}~\bibnamefont
  {Packard}}, \bibinfo {author} {\bibfnamefont {M.~A.}\ \bibnamefont {Bedau}},
  \bibinfo {author} {\bibfnamefont {A.}~\bibnamefont {Channon}}, \bibinfo
  {author} {\bibfnamefont {T.}~\bibnamefont {Ikegami}}, \bibinfo {author}
  {\bibfnamefont {S.}~\bibnamefont {Rasmussen}}, \bibinfo {author}
  {\bibfnamefont {K.~O.}\ \bibnamefont {Stanley}}, \ and\ \bibinfo {author}
  {\bibfnamefont {T.}~\bibnamefont {Taylor}},\ }\href {\doibase
  10.1162/artl_a_00291} {\bibfield  {journal} {\bibinfo  {journal} {Artificial
  Life}\ }\textbf {\bibinfo {volume} {25}},\ \bibinfo {pages} {93} (\bibinfo
  {year} {2019})},\ \Eprint
  {http://arxiv.org/abs/https://direct.mit.edu/artl/article-pdf/25/2/93/1896734/artl\_a\_00291.pdf}
  {https://direct.mit.edu/artl/article-pdf/25/2/93/1896734/artl\_a\_00291.pdf}
  \BibitemShut {NoStop}%
\bibitem [{\citenamefont {Dolson}\ \emph {et~al.}(2019)\citenamefont {Dolson},
  \citenamefont {Vostinar}, \citenamefont {Wiser},\ and\ \citenamefont
  {Ofria}}]{dolsoninno}%
  \BibitemOpen
  \bibfield  {author} {\bibinfo {author} {\bibfnamefont {E.}~\bibnamefont
  {Dolson}}, \bibinfo {author} {\bibfnamefont {A.}~\bibnamefont {Vostinar}},
  \bibinfo {author} {\bibfnamefont {M.}~\bibnamefont {Wiser}}, \ and\ \bibinfo
  {author} {\bibfnamefont {C.}~\bibnamefont {Ofria}},\ }\href {\doibase
  10.1162/artl_a_00280} {\bibfield  {journal} {\bibinfo  {journal} {Artificial
  Life}\ }\textbf {\bibinfo {volume} {25}},\ \bibinfo {pages} {50} (\bibinfo
  {year} {2019})}\BibitemShut {NoStop}%
\bibitem [{\citenamefont {Chalmers}(2006)}]{Chalmers2006-CHASAW}%
  \BibitemOpen
  \bibfield  {author} {\bibinfo {author} {\bibfnamefont {D.~J.}\ \bibnamefont
  {Chalmers}},\ }in\ \href@noop {} {\emph {\bibinfo {booktitle} {The
  Re-Emergence of Emergence: The Emergentist Hypothesis From Science to
  Religion}}},\ \bibinfo {editor} {edited by\ \bibinfo {editor} {\bibfnamefont
  {P.}~\bibnamefont {Davies}}\ and\ \bibinfo {editor} {\bibfnamefont
  {P.}~\bibnamefont {Clayton}}}\ (\bibinfo  {publisher} {Oxford University
  Press},\ \bibinfo {year} {2006})\BibitemShut {NoStop}%
\bibitem [{\citenamefont {Bedau}\ and\ \citenamefont
  {Humphreys}(2008)}]{Bedau2008EmergenceC}%
  \BibitemOpen
  \bibfield  {author} {\bibinfo {author} {\bibfnamefont {M.}~\bibnamefont
  {Bedau}}\ and\ \bibinfo {author} {\bibfnamefont {P.}~\bibnamefont
  {Humphreys}}\ }(\bibinfo {year} {2008})\BibitemShut {NoStop}%
\bibitem [{\citenamefont
  {Bedau}(1997)}]{https://doi.org/10.1111/0029-4624.31.s11.17}%
  \BibitemOpen
  \bibfield  {author} {\bibinfo {author} {\bibfnamefont {M.~A.}\ \bibnamefont
  {Bedau}},\ }\href {\doibase https://doi.org/10.1111/0029-4624.31.s11.17}
  {\bibfield  {journal} {\bibinfo  {journal} {Noûs}\ }\textbf {\bibinfo
  {volume} {31}},\ \bibinfo {pages} {375} (\bibinfo {year} {1997})},\ \Eprint
  {http://arxiv.org/abs/https://onlinelibrary.wiley.com/doi/pdf/10.1111/0029-4624.31.s11.17}
  {https://onlinelibrary.wiley.com/doi/pdf/10.1111/0029-4624.31.s11.17}
  \BibitemShut {NoStop}%
\bibitem [{10.(2021)}]{10.1162/isal_a_00382}%
  \BibitemOpen
  \href {\doibase 10.1162/isal_a_00382} {\emph {\bibinfo {title} {{A
  graph-theoretic approach to understanding emergent behavior in physical
  systems}}}},\ \bibinfo {series} {ALIFE 2021: The 2021 Conference on
  Artificial Life}, Vol.\ \bibinfo {volume} {ALIFE 2021: The 2021 Conference on
  Artificial Life}\ (\bibinfo {year} {2021})\ \bibinfo {note} {63},\ \Eprint
  {http://arxiv.org/abs/https://direct.mit.edu/isal/proceedings-pdf/isal/33/63/1929977/isal\_a\_00382.pdf}
  {https://direct.mit.edu/isal/proceedings-pdf/isal/33/63/1929977/isal\_a\_00382.pdf}
  \BibitemShut {NoStop}%
\bibitem [{\citenamefont {Ellis}(2011)}]{ellis2011top}%
  \BibitemOpen
  \bibfield  {author} {\bibinfo {author} {\bibfnamefont {G.~F.}\ \bibnamefont
  {Ellis}},\ }\href@noop {} {\bibfield  {journal} {\bibinfo  {journal}
  {Interface Focus}\ } (\bibinfo {year} {2011})}\BibitemShut {NoStop}%
\bibitem [{\citenamefont {Walker}\ \emph {et~al.}(2012)\citenamefont {Walker},
  \citenamefont {Cisneros},\ and\ \citenamefont
  {Davies}}]{walker2012evolutionary}%
  \BibitemOpen
  \bibfield  {author} {\bibinfo {author} {\bibfnamefont {S.~I.}\ \bibnamefont
  {Walker}}, \bibinfo {author} {\bibfnamefont {L.}~\bibnamefont {Cisneros}}, \
  and\ \bibinfo {author} {\bibfnamefont {P.~C.}\ \bibnamefont {Davies}},\
  }\href@noop {} {\bibfield  {journal} {\bibinfo  {journal} {Proceedings of
  Artificial Life XIII}\ ,\ \bibinfo {pages} {283}} (\bibinfo {year}
  {2012})}\BibitemShut {NoStop}%
\bibitem [{\citenamefont {Walker}(2014)}]{topdown}%
  \BibitemOpen
  \bibfield  {author} {\bibinfo {author} {\bibfnamefont {S.~I.}\ \bibnamefont
  {Walker}},\ }\href@noop {} {\bibfield  {journal} {\bibinfo  {journal}
  {Information}\ }\textbf {\bibinfo {volume} {5}},\ \bibinfo {pages} {424}
  (\bibinfo {year} {2014})}\BibitemShut {NoStop}%
\bibitem [{\citenamefont {Adams}\ \emph
  {et~al.}(2017{\natexlab{a}})\citenamefont {Adams}, \citenamefont {Zenil},
  \citenamefont {Davies},\ and\ \citenamefont {Walker}}]{me2017}%
  \BibitemOpen
  \bibfield  {author} {\bibinfo {author} {\bibfnamefont {A.}~\bibnamefont
  {Adams}}, \bibinfo {author} {\bibfnamefont {H.}~\bibnamefont {Zenil}},
  \bibinfo {author} {\bibfnamefont {P.}~\bibnamefont {Davies}}, \ and\ \bibinfo
  {author} {\bibfnamefont {S.}~\bibnamefont {Walker}},\ }\href@noop {}
  {\bibfield  {journal} {\bibinfo  {journal} {Scientific Reports}\ }\textbf
  {\bibinfo {volume} {7}},\ \bibinfo {pages} {997} (\bibinfo {year}
  {2017}{\natexlab{a}})}\BibitemShut {NoStop}%
\bibitem [{\citenamefont {Adams}\ \emph
  {et~al.}(2017{\natexlab{b}})\citenamefont {Adams}, \citenamefont {Berner},
  \citenamefont {Davies},\ and\ \citenamefont {Walker}}]{e19090461}%
  \BibitemOpen
  \bibfield  {author} {\bibinfo {author} {\bibfnamefont {A.~M.}\ \bibnamefont
  {Adams}}, \bibinfo {author} {\bibfnamefont {A.}~\bibnamefont {Berner}},
  \bibinfo {author} {\bibfnamefont {P.~C.~W.}\ \bibnamefont {Davies}}, \ and\
  \bibinfo {author} {\bibfnamefont {S.~I.}\ \bibnamefont {Walker}},\ }\href
  {\doibase 10.3390/e19090461} {\bibfield  {journal} {\bibinfo  {journal}
  {Entropy}\ }\textbf {\bibinfo {volume} {19}} (\bibinfo {year}
  {2017}{\natexlab{b}}),\ 10.3390/e19090461}\BibitemShut {NoStop}%
\bibitem [{\citenamefont {Zenil}\ \emph {et~al.}(2018)\citenamefont {Zenil},
  \citenamefont {Hernández-Orozco}, \citenamefont {Kiani}, \citenamefont
  {Soler-Toscano}, \citenamefont {Rueda-Toicen},\ and\ \citenamefont
  {Tegnér}}]{hectormain}%
  \BibitemOpen
  \bibfield  {author} {\bibinfo {author} {\bibfnamefont {H.}~\bibnamefont
  {Zenil}}, \bibinfo {author} {\bibfnamefont {S.}~\bibnamefont
  {Hernández-Orozco}}, \bibinfo {author} {\bibfnamefont {N.}~\bibnamefont
  {Kiani}}, \bibinfo {author} {\bibfnamefont {F.}~\bibnamefont
  {Soler-Toscano}}, \bibinfo {author} {\bibfnamefont {A.}~\bibnamefont
  {Rueda-Toicen}}, \ and\ \bibinfo {author} {\bibfnamefont {J.}~\bibnamefont
  {Tegnér}},\ }\href@noop {} {\bibfield  {journal} {\bibinfo  {journal}
  {Entropy}\ }\textbf {\bibinfo {volume} {20}},\ \bibinfo {pages} {605}
  (\bibinfo {year} {2018})}\BibitemShut {NoStop}%
\bibitem [{\citenamefont {Bedau}\ and\ \citenamefont
  {Packard}(1992)}]{packard1}%
  \BibitemOpen
  \bibfield  {author} {\bibinfo {author} {\bibfnamefont {M.~A.}\ \bibnamefont
  {Bedau}}\ and\ \bibinfo {author} {\bibfnamefont {N.~H.}\ \bibnamefont
  {Packard}},\ }in\ \href@noop {} {\emph {\bibinfo {booktitle} {Artificial Life
  II, Santa Fe Institute Studies in the Sciences of Complexity}}},\ \bibinfo
  {editor} {edited by\ \bibinfo {editor} {\bibfnamefont {I.~C.}\ \bibnamefont
  {Langton}}, \bibinfo {editor} {\bibfnamefont {C.}~\bibnamefont {Taylor}},
  \bibinfo {editor} {\bibfnamefont {D.}~\bibnamefont {Farmer}}, \ and\ \bibinfo
  {editor} {\bibfnamefont {S.}~\bibnamefont {Rasmussen}}}\ (\bibinfo
  {publisher} {M. A. Bedau and N. H. Packard},\ \bibinfo {year}
  {1992})\BibitemShut {NoStop}%
\bibitem [{\citenamefont {Zenil}\ \emph {et~al.}(2015)\citenamefont {Zenil},
  \citenamefont {Soler-Toscano}, \citenamefont {Delahaye},\ and\ \citenamefont
  {Gauvrit}}]{hector2}%
  \BibitemOpen
  \bibfield  {author} {\bibinfo {author} {\bibfnamefont {H.}~\bibnamefont
  {Zenil}}, \bibinfo {author} {\bibfnamefont {F.}~\bibnamefont
  {Soler-Toscano}}, \bibinfo {author} {\bibfnamefont {J.-P.}\ \bibnamefont
  {Delahaye}}, \ and\ \bibinfo {author} {\bibfnamefont {N.}~\bibnamefont
  {Gauvrit}},\ }\href@noop {} {\bibfield  {journal} {\bibinfo  {journal} {PeerJ
  Computer Science}\ }\textbf {\bibinfo {volume} {1}} (\bibinfo {year}
  {2015})}\BibitemShut {NoStop}%
\end{thebibliography}%
\end{document}